\documentclass[10pt,conference,compsocconf]{IEEEtran}

\usepackage{cite}
\usepackage{amsmath,amsthm,amssymb,amsfonts}
\usepackage{algorithmic}
\usepackage{graphicx}
\usepackage{textcomp}
\usepackage{xcolor}
\usepackage{soul}
\usepackage{url}
\usepackage{numprint}
\usepackage[caption=false,font=footnotesize]{subfig}
\usepackage{comment} 
\usepackage{balance} 

\def\BibTeX{{\rm B\kern-.05em{\sc i\kern-.025em b}\kern-.08em
    T\kern-.1667em\lower.7ex\hbox{E}\kern-.125emX}}
    
\npstyleenglish
\newcommand{\monthyear}{April 2022}
\newcommand{\cryptomarketsize}{1.88 trillion }
\newcommand{\polkadotsize}{18 billion }

\makeatletter
\def\ps@IEEEtitlepagestyle{
  \def\@oddfoot{\authorcopy}
  \def\@evenfoot{}
}
\def\authorcopy{
  {\footnotesize
  \begin{minipage}{\textwidth}
  This is a personal copy of the authors. Not for redistribution. The final version of the paper will be available through the IEEExplore Digital Library, in the proceedings of the 5$^{th}$ IEEE International Conference on Blockchain, Espoo, Finland, August 22-25 2022.
  \end{minipage}
  }
}

\begin{document}

\title{Analysis of Polkadot: Architecture,  Internals,  and  Contradictions}

\author{\IEEEauthorblockN{Hanaa Abbas}
\IEEEauthorblockA{\textit{College of Science and Engineering} \\
\textit{Hamad Bin Khalifa University}\\
Doha, Qatar \\
haab09879@hbku.edu.qa}
\and
\IEEEauthorblockN{Maurantonio Caprolu}
\IEEEauthorblockA{\textit{College of Science and Engineering} \\
\textit{Hamad Bin Khalifa University}\\
Doha, Qatar \\
macaprolu@hbku.edu.qa}
\and
\IEEEauthorblockN{Roberto Di Pietro}
\IEEEauthorblockA{\textit{College of Science and Engineering} \\
\textit{Hamad Bin Khalifa University}\\
Doha, Qatar \\
rdipietro@hbku.edu.qa}}


\maketitle

\begin{abstract}
Polkadot is a network protocol launched in 2020 with the ambition of unlocking the full potential of blockchain technologies. Its novel  multi-chain protocol allows arbitrary data to be transferred across heterogeneous blockchains, enabling the implementation of a wide range of novel use cases.
The Polkadot architecture is based on the principles of \textit{sharding}, which promises to solve scalability and interoperability shortcomings that encumber many existing blockchain-based systems. Lured by these impressive features, investors immediately appreciated the Polkadot project, which is now firmly ranked among the top 10 cryptocurrencies by capitalization (around 20 Billions USD). However, Polkadot has not received the same level of attention from academia that other proposals in the crypto domain have received so far, like Bitcoin, Ethereum, and Algorand, to cite a few. Polkadot architecture is described and discussed only in the grey literature, and very little is known about its internals.\\
\noindent{}
In this paper, we provide the first systematic study on the Polkadot environment, detailing its protocols, governance, and economic model. Then, we identify several limitations---supported by an empirical analysis of its ledger---that could severely affect the scalability and overall security of the network. Finally, based on our analysis, we provide future directions to inspire researchers to investigate further the Polkadot ecosystem and its pitfalls in terms of performance, security, and network aspects.\\
\end{abstract}


\IEEEpeerreviewmaketitle

\section{Introduction}
\label{sec:introduction}
Interest in cryptocurrencies has been steadily rising ever since the introduction of Bitcoin and blockchain---the foundation of all existing crypotocurrencies---in 2009. As of \monthyear, the global cryptomarket capitalization is estimated at more than \cryptomarketsize US dollars. Not only did blockchain enable secure financial transactions, but it also laid the path for transforming cooperation over the Internet, allowing ``trusted cooperation between untrusted entities" \cite{guo2019graph}. As a technology of promising prospects, blockchain and its applications have continued to witness rapid growth in terms of development and public attraction over the years. 

This paper focuses on a relatively new cryptocurrency---Polkadot---whose network was only launched mid-May 2020, yet had successfully secured a spot amongst the top 10 cryptocurrencies, owning a market capitalization of around \polkadotsize USD  \cite{coinmarketcap}. Polkadot is developed by the Web3 Foundation and Parity Technologies, both founded by Dr. Gavin Wood, who is also known for being a co-founder and former CTO of Ethereum. In 2016, Dr. Gavin Wood shared his vision of Polkadot and its technical outline through the white paper \cite{wood2016polkadot}; which was, in his words, creating ``the next version of Ethereum'' that would complement the aspects for which Ethereum 2.0 is less optimal \cite{polkadotnetwork}.

Polkadot is a fully ``\textit{sharded}" blockchain whose design is based on sharding---a database splitting technique---that enables multiple chains to process their transactions in parallel. Each blockchain shard is called a ``\textit{parachain}" which connects to the \textit{Relay Chain}. The Relay Chain is effectively the heart of the blockchain, acting as the main \textit{hub} of the system. Furthermore, Polkadot serves as an interoperability platform; i.e., it allows cross-communication between heterogeneous blockchains including external ones, such as Bitcoin and Ethereum. While the Relay Chain orchestrates and ensures correct functioning of the entire network, parachains can be customized as needed, including to host smart contracts and a myriad of other use cases. Essentially, Polkadot was developed with the goal to address some shortcomings of existing blockchain technologies, namely, scalability, interoperability, and achieving standard security guarantees across differing trust models. In the rest of this paper, we will test whether these objectives have been achieved.

Due to the novelty of multi-chains, there is a definite need to investigate their design and processes, particularly within an active ecosystem. As one of the first publicly released multi-chain projects, Polkadot is a perfect case study for evaluating the security properties of cross-chain communications among heterogeneous blockchains.
However, despite achieving a good standing and wide consideration in the market, Polkadot has not received the same level of attention from academia enjoyed by  other proposals, such as Bitcoin, Ethereum, and Algorand, to cite a few. 
In fact, although well described in the grey literature, the Polkadot architecture and its practice is still little discussed in the scientific community. 
To fill this gap, we shed light on the complex ecosystem of Polkadot through a preliminary analysis of its architecture, followed by the identification of several contradictions that could limit the strengths envisaged by its creators. Our main goal is to propose the first systematic study on the Polkadot ecosystem, providing researchers interested in this field with a solid knowledge base to understand the Polkadot network and other blockchains based on multi-chain protocols. The highlighted limitations, calling for contributions to mitigate their effects, and the proposed research directions complete the work.\\ 

\noindent
{\bf Contributions.}
In this paper, we provide several contributions:
\begin{itemize}
\item We first study and analyze the overall architecture of Polkadot, ranging from the roles of its network participants, consensus schemes, and governance to its economic and security models. To the best of our knowledge, this is the first systematic study to investigate Polkadot's architecture and protocols.
    
\item Based on our analysis, we highlight the main limitations and contradictions of Polkadot, calling for further investigations and research to propose viable solutions addressing the uncovered limitations. 

\item We validate our assertions by providing an empirical analysis of the validators' election data that we collected from the Polkadot blockchain, starting from the introduction of network validators in June 2020 until April 2022.\\
\end{itemize}

\noindent
{\bf Paper Organization.}
The remainder of the paper is organized as follows. Section \ref{sec:related_work} presents an overview of multi-chain design approaches existing in the literature and other multi-chain projects publicly available with a market presence. Section \ref{sec:polkadot_internals} extensively reviews Polkadot from different perspectives, including its architecture, consensus protocol, governance, and security. Further, in Section \ref{sec:polkadot_limitations}, we discuss the limitations we found in the Polkadot ecosystem, validated by an empirical analysis of the ledger. We then conclude the study in Section \ref{sec:conclusion}, providing possible extensions and directions for future work. 

\section{Related Work}
\label{sec:related_work}
Despite its promising prospects, blockchain technology has encountered some fundamental limitations over the course of its development. Real-world applications demand for secure inter-communication between heterogeneous blockchain systems; for instance, for transferring  different assets, product lifecycle management \cite{9070689}, etc.. Several works (e.g., \cite{back2014enabling,10.1007/978-3-030-43725-1_3}) implemented two-way \textit{pegged sidechains} that enable bidirectional movement of assets and cross-chain communication between Proof of Work (PoW) sidechains, without the need for a trusted third party intermediary. Back et al. (2014) \cite{back2014enabling} implemented a two-way peg between Bitcoin and other sidechains, such that: (a) transfers are atomic; and, (b) there should be zero risk to the other party. 
In their implementation, a transaction is created on the parent chain to lock the assets, then another transaction on the sidechain generates the new asset, given valid proof of possession. 
Later, Kan et al. (2018) \cite{8431965} further developed the two-way peg architecture to enable inter-blockchain communication between heterogeneous systems. Besides achieving blockchain interoperability \cite{belchior2021survey}, other works relied on the multi-chain architecture for resolving scalability and throughput issues in traditional blockchains, e.g., for uses-cases in IoT \cite{9154979} and auditing applications \cite{8761448}.

To the best of our knowledge, there are a few multi-chain projects that both improve transactions' scalability and implement cross-chain communication. The first project is Cosmos \cite{cosmos}, which launched in March 2019 but did not approach the same level of market capitalization as Polkadot did. In Cosmos, there are multiple independent chains operating in `\textit{zones}' that enjoy trust-free inter-chain communication via the main \textit{hub} chain. The second is the anticipated version of Ethereum (formerly known as Ethereum 2.0); however, Ethereum 2.0 is still to be launched, approximately, later in 2023. As per Polkadot, to the best of our knowledge, there is no work in the literature that analyzes the proposed architecture, having in mind to check its true abiding to the principles of scalability, interoperability and decentralization. In this respect, our contribution is the first one to show that the limitations of the proposed architecture introduce some tensions with respect to the above discussed objectives. 

\begin{figure}[!t]
\includegraphics[width=\columnwidth]{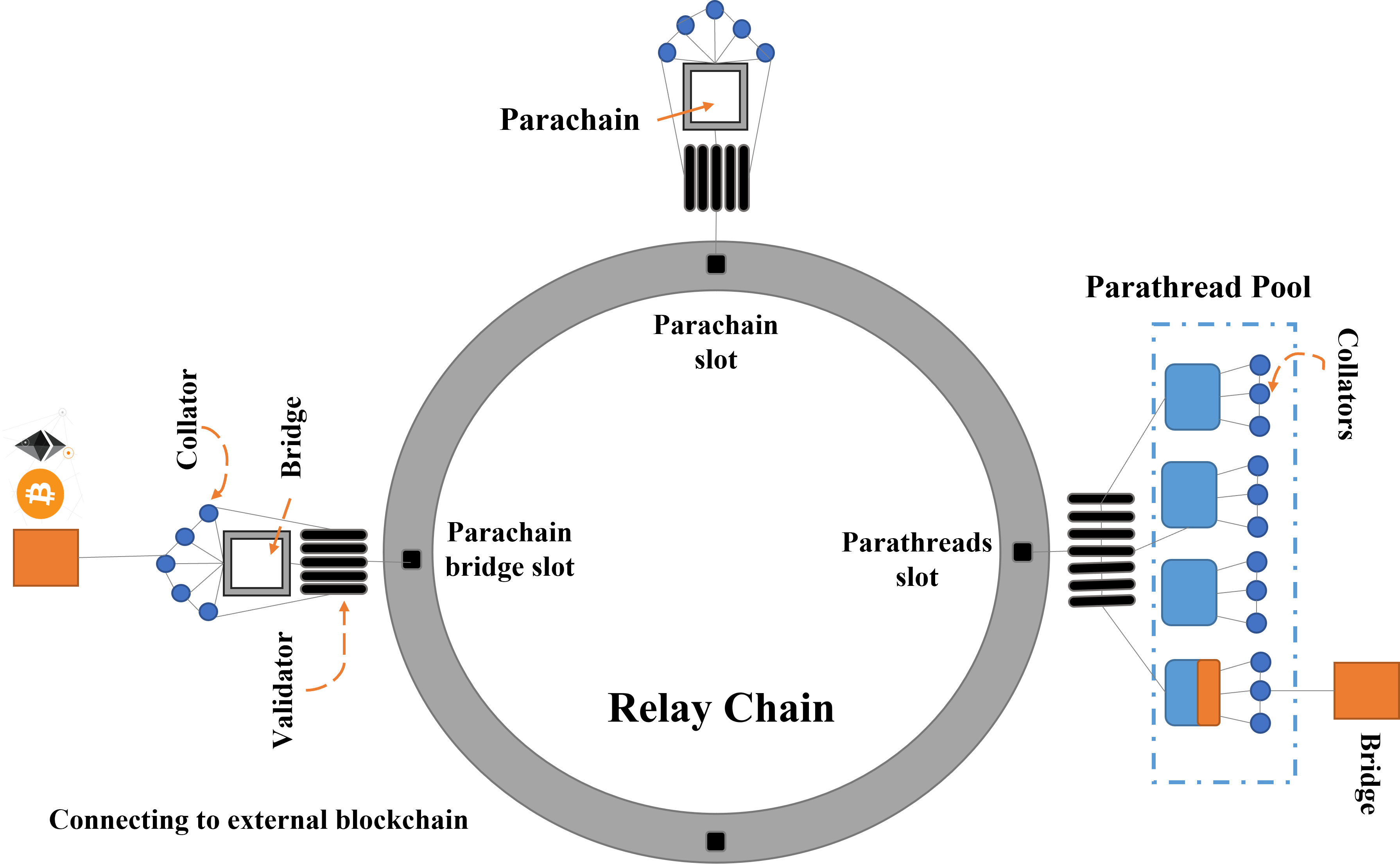} 
\caption[Polkadot Data Structures and Participants]{Polkadot Data Structures and Participants} 
\label{Fig:polkadot_full_architecture}
\end{figure}

\section{Polkadot Internals}
\label{sec:polkadot_internals}
Polkadot is popular for being a multi-chain (sharded) blockchain network, specifically built to support the integration of heterogeneous blockchains in a trust-free environment and under common security guarantees. It also enables secured communication with external sovereign blockchains (e.g., Bitcoin or Ethereum); thus, laying the foundations for establishing a truly decentralized web. In addition to interoperability, Polkadot's multi-chain design allows transactions to be processed in parallel instead of sequentially; leading to a higher throughput and scalability. All these properties are provided owing to the \textit{Relay Chain}. It unites a network of heterogeneous blockchains (or arbitrary state machines), operating in parallel, which include parachains, parathreads, and bridges. The Relay Chain provides shared security to the entire network through its consensus mechanisms. 
 
Polkadot adopts Proof of Stake (PoS) consensus, namely a variation of it called the \textit{Nominated Proof of Stake (NPoS)}. The scheme is referred to as ``nominated" because the \textit{validators} (Relay Chain maintainers: nodes that are responsible for authoring and verifying new blocks) are backed up by \textit{nominators} who lock their DOT --- Polkadot's native currency --- as collateral in exchange for staking rewards. However, if validators misbehave, their corresponding nominators are also slashed (i.e., DOT deduction from their accounts). While validators maintain the Relay Chain, a collator is a node responsible for maintaining its parachain. All previously described structures and roles are shown in Figure \ref{Fig:polkadot_full_architecture}.

\subsection{Paraobjects}
\label{sec:paraobjects}
In this subsection, we review the paraobjects supported by Polkadot, namely: parachains, parathreads, and bridges. 

\subsubsection{Parachains}
These are independent chains that operate in parallel, maintained by collators, and can communicate with each other via the Cross-Consensus Message Passing (XCMP) format. Parachains can have customized runtime logic, which includes project-specific rules for governance, economics and incentive mechanisms. The key constraint on parachains is that their state transition function (STF), which is stored on the Relay Chain, must be verifiable by validators. Proofs of new state transitions (aka, \textit{Proof of Validity or PoV}) that occur on a parachain are produced by collators and then validated against the registered STF by validators. Polkadot is designed to support a maximum of 100 parachain slots, where a slot can be acquired through governance (for \textit{common good} parachains) or via candle auctions. 

Common Good Parachains offer functionality that benefits the network as a whole. Since they are not allocated via the auctioning process, their slot lease does not expire and can only be revoked by the governance body. Examples of public utility chains include: asset transfer (fungible and non-fungible) chains, smart contract chains, and bridges.  

\textbf{Slot (Candle) Auctions.} In parachain slot auctions, parachain accounts lock their DOT as bids to secure a slot on the network for a specified lease duration. The lease duration is in three months increments up to a maximum of 2 years. Throughout the auction's ending period, a per-block snapshot is taken to capture the bids and winners status. From the snapshots, the `ending block' from which the winners are announced is randomly selected. This is a security feature to prevent malicious bidders from `sniping' the auction---waiting out until the auction is about to terminate to place the highest bid.
  
Parachain projects can bid with their own funds or source them using the \textbf{crowdloan} functionality, which is asking the community to stake DOT in support of their chain. The reserved DOT contributes to the security of the system just as staking does. Since parachain slots are scarce, placing sufficiently high bids and getting support from the community ensures that honest projects get connected. \\

\begin{figure}[!t]
\includegraphics[width=\linewidth]{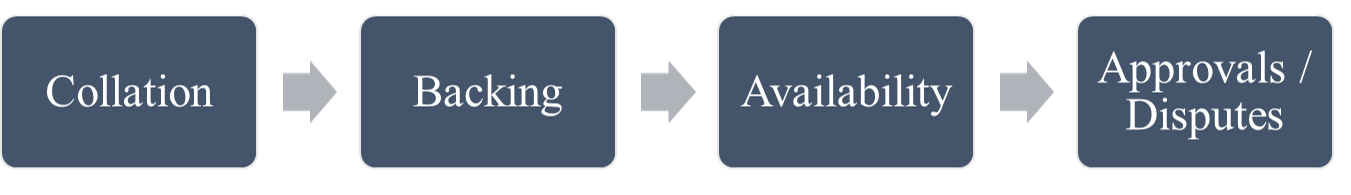}
\centering
\caption{Stages of Parachain Consensus}
\label{Fig:parachain_consensus}
\end{figure}

\noindent\textbf{Parachain Consensus.} Parachain consensus is the process of ensuring that only valid candidate blocks get stored on the Relay Chain. Figure \ref{Fig:parachain_consensus} lists the phases of parachain consensus. The first process, collation, is performed on the parachain by collators, where they produce parachain blocks and send them to the assigned set of para-validators for the next phase, backing. Backing is the process where the validators conduct initial validity checks on the submitted parachain blocks, and if the checks pass, they dispatch ``candidate receipts" to the Relay Chain transactions pool, where further validation and consensus decide if the candidate would be included in the Relay Chain block. \textit{Availability} scheme ensures that the necessary data is available to all validators so that they are able to validate the parachain block if needed. In the final stage, validators indicate their approval or disputes about the block by voting, and the super-majority vote ($2/3f + 1$) wins. 

\noindent
The four-staged process is explained hereafter:\\
\noindent
\textbf{Collation.} Collators aggregate all transactions created on the parachain and produce a candidate block. Along with the proposed block, the collator provides PoV block which is a summary of the parachain's state transition.\\
\noindent
\textbf{Backing.} After a para-validator receives the PoV block, it validates it against the parachain's registered STF. If the block passes the checks, the validator gossips it to the remaining para-validators who also attest its validity. When majority of the para-validators agree on the block, they sign the PoV's header and prepare a candidate receipt. The candidate receipt is gossiped to the other Relay Chain validators. After which, a subset of other validators also validate the PoV block and submit their attestations of validity. If the verification fails at any stage, the block is immediately rejected.\\ 
\noindent
\textbf{Availability.} In addition, the availability of the parachain blocks is checked by Relay Chain validators. The parachain blob (PoV block and set of outgoing XCMP messages) need to be available for some time to all validators for the finality (see Section \ref{grandpa}) process. The para-validators construct erasure coding chunks of the parachain blob and gossip a piece to each validator. Enough validators must report that they posses an erasure code piece to pass the availability test. At least $1/3 + 1$ of the code pieces are needed to construct the full block whenever needed.\\
\noindent
\textbf{Approvals/Disputes.} These are collected on the Relay Chain as part of the block finalization process (GRANDPA consensus). If the block fails any of the previous checks, the corresponding para-validators/offenders are slashed.\\

\subsubsection{Parathreads} 
Unlike parachains, parathreads do not lease a dedicated slot on the Relay Chain. Instead, they form a pool together that shares a designated number of slots. In other words, the only difference between parachains and parathreads is economic. In order for a parathread to submit its block candidate to the Relay Chain, it must participate in a block-by-block auction competing with other parathreads. Collators signal to the assigned para-validators their bid in DOT, and the block author selects the candidates with the highest bids to include on the Relay Chain. This pay-as-you-go model makes parathreads suitable for applications that do not require high throughput or run less frequently. Parathreads are also an alternative for projects that cannot afford to lease a parachain slot.

\subsubsection{Bridges}
A Polkadot bridge is a specialized parachain, also maintained by collators, that handles the communication between the Relay Chain and an external sovereign blockchain (e.g., Bitcoin). In this respect, bridges offer two-way compatibility and interoperability between blockchains; allowing arbitrary data exchange between two networks.  

\subsection{Consensus Roles}
\label{sec:roles}
\begin{figure}[!t]
\includegraphics[scale=0.45]{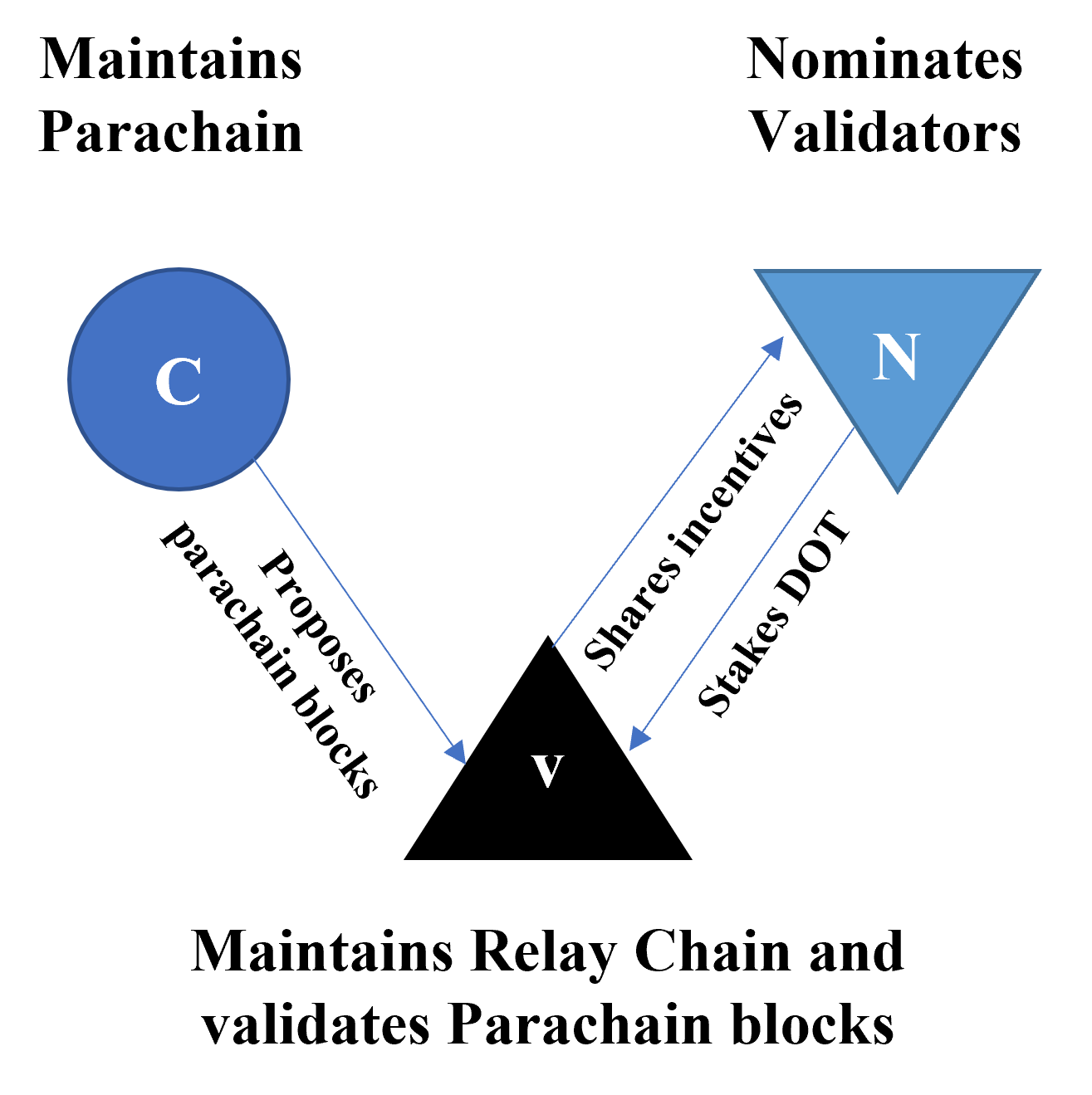}
\centering
\caption{Interaction between Polkadot roles} 
\label{Fig:polkadot_roles}
\end{figure}

In this section, we present the network maintainers, validators and collators, and nominators---the entities that  contribute to the NPoS consensus. Collators and validators participate in the network by running a full (highly available) node that stores the network's data, however nominators rely on a light client that retrieves only specific validator-related data. The interactions between the roles of Polkadot participants are depicted in Figure \ref{Fig:polkadot_roles}, while Table~\ref{tab:scalability_limitations} shows the limitations set by the protocol on the number of participants/paraobjects.

\subsubsection{Validators}
Polkadot restricts the number of active validators on the network. As the number of connected parachains increases, the active set will reach the maximum size of \numprint{1000} validators. This limit is the result of a compromise between security and performance. The larger the number of validators in the system, the higher its security; however, the throughput of the system is reduced since more time is needed to reach consensus between all the nodes.   

Validators are backed by self-stake and by nominators who bond a sufficiently large amount of DOT in their favor. This NPoS model contributes economic security to the network.
Since majority of the actors are assumed to be honest, and at most a third can be malicious, validators are incentivized to behave honestly because malicious activities, if detected, are heavily punished in the form of major slashing of their staked funds. Since slashing applies to the total stake of a validator, a validator bonding a large self-stake would communicate higher confidence and trust to potential nominators.

Validators are the maintainers of the Relay Chain. They engage in several tasks in parallel that include: proposing new relay chain blocks, reaching consensus about the finalized chain, validating the state transition function of parachains, and ensuring parachains' data availability and validity. In exchange for infrastructure and operational costs involved in running its node, a validator can set a variable commission rate (0\%-100\%) --- which is a percentage of the total block rewards that the validator had earned in a completed era. Era points are gained when the validator issues validity statements for a parachain block and generates a valid relay chain block. The commission is initially deducted from the era rewards and the remaining rewards are paid to the validator and his nominators according to their stake distribution. Validators are also rewarded for authoring new blocks by receiving a portion (20\%) of the transaction fees and 100\% of the (optional) transaction tips. 

\subsubsection{Collators}
Collators are responsible for maintaining a full node on their parachain and a full node on the Relay Chain. Meaning, they can access all necessary information, including the Relay Chain's state transitions, needed to author new blocks. In summary, the collators tasks are to: (a) maintain their respective parachain and store its data; (b) exchange messages with other parachains using XCM format; (c) produce parachain block candidates; (d) generate parachain state transition proof, also known as PoV block, which is a bundled proof/summary of the parachain's state transitions; and, (e) in case of a parathread collator, offer a bid in DOT in the block-by-block auction.

\subsubsection{Nominators}
Nominators can nominate up to 16 (trusted) validators daily by bonding a part of their DOT as collateral (aka, staking). In return, nominators receive part of the block rewards that their active nominations earned in proportion to the stake they bonded. However, if the active validator commits a punishable offense (e.g., being unresponsive or equivocating), the nominator also gets slashed and loses a portion of his DOT. 

\textbf{Phragm\'{e}n Election.} Validators are elected into the active set based on the phragm\'{e}n election which implements proportionally justified representation (a property in voting theory). Phragmén's election has two major goals that are to: (a) select a subset of validators from a larger set based on stake-weighted votes; and, (b) spread out the stake backing each validator equally as much as possible. For further technical details about how the election algorithm works, refer to the research paper in \cite{polkadotwiki} and \cite{cevallos2021verifiably}.

\subsection{Consensus Mechanisms}
\label{sec:consensus}
Polkadot adopts a hybrid consensus mechanism where block production and finality are two isolated processes. The algorithm responsible for handling block production is called BABE (Section \ref{subsec:babe}), and the finality gadget is GRANDPA (Section \ref{grandpa}). We cover these algorithms in more detail later in this subsection. On the one hand, block production (BABE) is intended to be \textit{probabilistically safe} (able to continuously generate new blocks with a probabilistic assurance that the generated blocks will be finalized after some time) \cite{burdges2020overview}. On the other hand, GRANDPA provides deterministic and \textit{provable finality} (universal agreement on the canonical chain with zero probability of reversion except by a hard fork) \cite{burdges2020overview}. Provable finality is a desirable feature specifically for achieving secure interoperability with external blockchains; it provides guarantees that data stored on Polkadot or its parachains are final and have almost no chance of being easily reverted. By combining these two mechanisms, Polkadot allows block production to continue rapidly, while the slower finality mechanism runs asynchronously without risking slower processing times or stalling. In what follows, we describe the NPoS and the hybrid consensus scheme adopted by Polkadot. 

\begin{table}
\centering
\caption{Polkadot's Limitations on the Number of Participants/Paraobjects}
\label{tab:scalability_limitations}
\begin{tabular}{|l|l|} 
\hline
\textbf{Entity} & \textbf{Limit} \\ 
\hline
Validators & 297 (max. 1000) \\ 
\hline
Nominators & 22,500 \\ 
\hline
Parachains & 14 (max. 100) \\ 
\hline
Council & 13 (max. 24) \\
\hline
\end{tabular}
\end{table}

\subsubsection{NPoS (Nominated Proof of Stake)}
In NPoS, a large number of nominators back a limited set of validators. NPoS is comparable to Delegated Proof of Stake (DPoS); however, the key difference between them is that in NPoS, nominators are subject to slashing if their validator is malicious, whereas delegators (or witnesses) are not. As the nominators share slashings as well as economic benefits, they are incentivized to choose honest validators with a record of good performance. Thus, the NPoS scheme is expected to contribute to the overall economic security of the network.

\subsubsection{BABE (Blind Assignment for Blockchain Extension)}
\label{subsec:babe}
BABE is a slot-based block production algorithm. Time is represented as epochs (4 hours), each consisting of 2400 slots, i.e., each slot is 6 seconds long. In Polkadot, the target block time is also 6 seconds; meaning that within each slot, a new block should be produced and appended to the Relay Chain. BABE is responsible for orchestrating the block production process, which handles the following mechanisms: (a) selection of block producers; (b) selection of the canonical chain; and, (c) adjustment of slot time. We describe these three processes hereafter. 

\textbf{Selection of Block Producers.} In each slot, BABE assigns primary author(s) and a secondary author to produce a Relay Chain block. Primary slot leaders are selected randomly based on a Verifiable Random Function (VRF). The inputs of the VRF are: epoch randomness seed, slot number, and author's private VRF key. Because of the randomness in the selection process, some slots may end up with more than one primary producer and, sometimes even, a slot may have zero primary producers. To handle the case of no primary producers, BABE uses a fall-back round-robin algorithm to assign secondary slot leaders. In each slot, a secondary producer is selected and produces a block along with the primary producer(s), if any. However, if a valid primary block is produced, the secondary one is ignored.   

When multiple validators are primary producers in a given slot, all of them produce a block and broadcast it in the network, and the block that reaches most of the network first wins the race. Due to network conditions, forks may emerge until the finalization algorithm (GRANDPA) selects the canonical chain and kills other forks. 

At the beginning of an epoch, each validator evaluates its VRF for each slot in the entire epoch's duration. If the result of the VRF is below a threshold value defined in the protocol, the validator knows that it is a primary author for that slot. However, other participants in the network are not aware of that assignment in advance (hence the name `Blind Assignment'). Not knowing the assignment beforehand adds to the security of the system; unlike round-robin algorithms, adversaries cannot launch a coordinated attack against the next author.
 
\textbf{Selection of the Canonical Chain.} There are two rules for selecting the best chain in BABE. First, block producers must build on the latest block that has been finalized by GRANDPA. Second, when there are forks after the finalized head, BABE builds on the chain with the most primary blocks. 

\textbf{Adjustment of Slot Time.} BABE does not rely on a centralized clock authority for time synchronization. Instead, it aligns slot numbers using a relative time protocol. When a validator node receives a block, it checks the block's timestamp and slot number, compares it to its local time, and forecasts future slot times. More information about local time synchronization in BABE is provided in \cite{burdges2020overview} and \cite{babe_web3}. 

\subsubsection{GRANDPA---Finality Gadget}
\label{grandpa}
GRANDPA stands for GHOST-based Recursive ANcestor Deriving Prefix Agreement. It allows a set of nodes to agree on the canonical chain out of many possible forks. It works under the assumption of a Byzantine Fault Tolerant system: a partially synchronous network with at most one third Byzantine (dishonest or unresponsive) nodes. GRANDPA reaches agreements on chains rather than blocks; in other words, whenever more than two thirds of validators vote on a chain containing a certain block, all blocks up to that one are finalized at once. For more details about GRANDPA, refer to \cite{burdges2020overview} and \cite{stewart2020grandpa}. 

\subsection{Security Model}
\label{sec:security_model}
The adversarial model of Polkadot assumes that at least two thirds of the active validators are honest. The model also assumes that parachains are untrusted clients of the Relay Chain; and hence, there are no specific security assumptions enforced on the parachains or their collators. Nonetheless, Polkadot assumes that every parachain has one honest and reachable collator \cite{burdges2020overview}. 

Security is shared in Polkadot---all connected parachains automatically benefit from same level of security provided by the Relay Chain. This is to ensure the validity of the entire system and that no individual component is corruptible. Security is independent of the number of parachains; it only depends on the validators who have a sufficiently large amount of DOT backing them. Nominators and validators contribute to the security of the network by locking funds, aka, staking. Nominators share in the rewards, but also the punishment, of the validators they vote for. Phragm\'{e}n’s election process picks the winners by finding the combination that leads to having the most value at stake. In PoS systems, the biggest indicators of the network's security are the value at stake and value at risk. Meaning, the greater the amount of DOT staked by honest validators and nominators, the higher the amount of DOT an attacker needs to acquire a validator slot.
   
The implication of a shared/pooled security is that, in order to revert a parachain’s block, an attacker would have to revert the entire Polkadot system, including all other parachains. This is due to the Relay Chain blocks storing proofs of validity from parachains, meaning that when the chain is finalized, that parachain block is also finalized. This shared security provides the necessary guarantees to enable trusted cross-chain communication between untrusted entities.

\subsection{Economic Model}
\label{sec:economic_model}
In this subsection, we describe Polkadot's economics and incentive layers.

\subsubsection{Transaction Fees}
Polkadot utilizes weight-based fees that are charged from the sender's account before the transaction is dispatched. Transaction fees are the sum of a weight fee and a length fee. The block weight represents the time it takes to produce and validate the block. Hence, each transaction has a base weight that accounts for its inclusion overhead (e.g., time required to verify sender's signature) and a dispatch weight for the execution time. The length fee is the size of the transaction in bytes multiplied by a per-byte fee. Moreover, the weight fee is adjustable according to current network conditions. 
A portion of the fee (20\%) goes to the block producer and the remainder goes to the treasury. Therefore, producers are likely to prioritize transactions with the highest fees for maximum rewards. Users can provide an optional tip in order to increase the transaction's priority, especially during congested periods. 

\subsubsection{Staking}
The staking system calculates block rewards for every era based on the total era points earned by validators. Points do not have a corresponding DOT value until the end of the era. Validators accumulate points based on their performance, primarily for signing validity statements and producing valid blocks.
Era points are expected to average out amongst validators, hence rewards are paid out almost equally (with slight variations) regardless of validators' total stake. The slight variations in payouts are due to probabilistic factors (randomness) in the block production and parachain validation assignments. 

After the commission payment is deducted, the staking rewards are distributed to the validator and nominators in proportion to their contributed stake. Thereby, the scheme provides incentives for more nominators to vote for lower-staked validators. This is because, the lower the validator's total stake is, the higher the percentage of the nominator's stake, and therefore the higher the rewards the nominator receives. Note that only the top 256 nominators receive staking rewards for an oversubscribed validator.

\subsubsection{Slashing}
If a validator misbehaves, punishment will be enforced on him and his nominators by losing a percentage of their staked DOT. Slashed DOT are added to the treasury, rather than being burned, since slashes can be reverted by the Council by simply returning the DOT from the treasury to the slashed accounts, given that the slash is protested and found unreasonable (e.g., a faulty runtime) within 28 days. Generally, slashing depends on the number of repeated offenses and how many other validators committed the same offense during the epoch. Meaning, slashing is super-linear wherein as the number of offenders increases, the slash percentage also increases.  

There are several degrees of offense where the percentage of stake deduction is decided accordingly. A validator is required to be available and to prove its availability; every session the node will send an "I'm Online" heartbeat. If a validator fails to produce any block during the epoch and send the heartbeat, it is declared unresponsive and might receive slashing. Slashing in the case of unresponsiveness depends on the number of repeated offenses and how many other validators were offline during the same epoch. 

A more serious offense is equivocation --- providing conflicting information to the network. For example, a validator signs multiple votes in the same GRANDPA round for different forks that conflict with each other, or a validator produces two or more blocks on the Relay Chain in the same BABE time slot. Repeated or concurrent equivocation is unlikely to be accidental, and therefore is severely slashed. If equivocations are detected from different validators, it is likely to be a coordinated attack. Offending validators will be kicked immediately and will lose their nominators. 

Some misconduct can be slashed up to 100\% of the stake because they pose a grave security or monetary risk or result in major collusion. An example would be casting GRANDPA votes for a chain that conflicts with an already finalized chain. The behavior is considered a definite attack because it is trying to revert finalized blocks.

\subsubsection{Inflation}
Polkadot has no maximum limit for DOT supply; hence, the network follows an inflation model. Block rewards are maximized when the staking rate (total amount staked over the total token supply) is at 50\% . The remainder of the rewards due to non-ideal staking rate goes to treasury.

\subsubsection{Accounts}
An account's address in Polkadot is represented by the user's public key which is 32-byte (256-bit) long. Polkadot uses SS58 address encoding format which allows account portability among compatible Substrate-based blockchains. Different chains have a different version prefix that is used to identify whether an account belongs to its chain or another. Consequently, a user needs only to generate a keypair once and use it on all chains, given that only the address's format would change from one chain to the other. Although this practice is discouraged for security reasons, it can be beneficial for specific use cases, e.g., when setting up a crowdloan account. 

Nominators and validators are recommended to set up two separate accounts, namely the Stash and Controller accounts. The controller account acts on behalf of the stash account, initiating nominating and validating actions. It is used to set staking preferences and only needs to contain enough DOT to pay transaction fees. Whereas, the stash account holds the actual funds bonded for staking. This keypair separation ensures higher protection for the accounts. 

\subsection{Governance}
\label{sec:governance}
Polkadot's governance constitutes of three entities: public referenda, council, and technical committee. Any DOT holder in Polkadot has the power to submit or endorse a proposal, vote on active referenda, nominate for a seat on the council, or vote for other council candidates. Council members are elected candidates whose roles include proposing, voting and vetoing public referenda. The technical committee is composed of the developer teams who are actively building Polkadot, e.g., Parity Technologies and Web3 Foundation. They are given the power to propose an emergency referendum. 

\subsubsection{Public Referenda Chamber}
Any DOT holder in the system can bond tokens to create or second a public referendum, where the voters also deposit an amount equal to the original bond. Every proposal is compiled and its preimage must be submitted to the blockchain. There can be a maximum of 100 public proposals in the proposal queue, which is made up of community proposals and council proposals. Every launch period (28 days), proposals with the highest bond will be moved to the referendum timetable.

The referendum to be voted upon alternates between the top proposal in either of the community or council queues. When a referendum is selected, the voting period (28 days) starts. Votes can be either `yes' or `no', or completely abstaining from voting. After the voting period, the votes are tallied which decide whether the referendum passed or not.

\subsubsection{Council} 
The council is currently composed of 13 members including Gavin Wood, the founder of Polkadot. In addition, the network plans to accommodate only a maximum of 24 seats on the council \cite{polkadotwiki}. Council members have a term of 7 days after which a new Phragm\'{e}n election takes place, and members are selected based on the stake backing them. Council members have two main tasks: proposing referenda and cancelling malicious referenda.

\subsubsection{Treasury}
Any DOT holder in the system can bond tokens to create a treasury proposal suggesting a spend from it. If the treasury proposal is declined by the council, 1\% of the tokens that were bonded to create the proposal are burned. The treasury can release funds only every 24 days.

\section{Polkadot Contradictions}
\label{sec:polkadot_limitations}
For a long time, Polkadot has maintained a steady position in terms of coin market capitalization, ranking amongst the top 10, meanwhile its network and architecture are still developing and not fully mature yet. In addition, even though Polkadot was introduced as a multi-chain interoperability platform, parachains were only added in November 2021, a year and half after the network's launch, and it is not known yet when bridges will become available to fulfill the interoperability claims. In the following subsections, we recap from our perspective numerous contradictions in the architecture and design of Polkadot, our empirical analysis, and discussion of the results. 

\subsection{Polkadot Limitations}
\noindent\textbf{Difficulties in becoming a validator.}
If you want to become a validator on the Polkadot network, you will have to overcome some hefty hurdles. The network is built to support a maximum of \numprint{1000} validators. Currently, only 297 validators are allowed into the active set per era (day); and, this number will slowly scale up to \numprint{1000} as the number of connected parachains increases. Meaning, validators must compete against one another to enter the active set on a daily basis, and only those with the highest stake, i.e., self-stake and nominators stake combined, win. This competition effectively forces a high minimum stake requirement on validators, making it extremely challenging for new joiners to participate and compete. Also, the collective stake is locked in the validator/nominators accounts, i.e., it is not transferable until an account decides to un-nominate or un-validate. In this case, 
the bonded amount remains locked for a further 28 days until it can be withdrawn. This may create a fertile ground for only a handful of big corporations who can invest such large amounts to become a validator, potentially leading to some level of centralized operation in the future.  \\   

\noindent\textbf{Problems with becoming a nominator.} Nominators face even greater obstacles. This is because validators can charge from 0 - 100\% commission and can change the rate at any time without prior notice to nominators. If a validator sets his commission to 100\%, none of the nominators get staking rewards for the duration of the session. Consequently, validators who charge lesser commission are more likely to be nominated, possibly causing them to become oversubscribed. However, only the top 256 nominators, for an oversubscribed validator, receive staking rewards and the rest lose on their bonded stake. Moreover, if a validator misbehaves or goes offline, all nominators who staked him, including those not eligible for rewards, will be slashed. Whist the block rewards are equally distributed to all active validators, slashing removes a percentage of the validators' total stake. For validators who set aside the minimal self-stake, the nominators are at greater risk; that is, they could lose more to slashing than what they would earn from the staking rewards. And, the slash could go up to 100\%, depending on the type of offence. 

Another major scalability concern is related to the validators phragm\'{e}n election. When the number of nominators increased beyond \numprint{30000}, a number of problems occurred on the Polkadot network: the block production time slowed down and the network suffered high instability. According to the statement released by Polkadot \cite{polkadot_postmortem}, its runtime crashed on May 24, 2021, with an out of memory error while trying to build a block containing the large election mapping results, due to the sizeable nominators set. At that time, validators were asked to downgrade their nodes' runtime version while Polkadot worked on a quick workaround (increasing the Wasm memory bounds). To this day, the problem persists, including on Kusama (Polkadot's canary network) where the nominators set was reduced to around \numprint{7300} participants as the number of validators reached the maximum \numprint{1000}.  

To address the aforementioned issue, Polkadot's governance approved a temporary solution---unfortunately, at the disadvantage of the nominators, as explained in the following. First, the nominators set is now limited to \numprint{22500} members. Second, in order to nominate, nominators must bond at least 120 DOT, which equates to \$2,160 at \$18/DOT. These restrictions come in addition to the original limit of allowing an account to nominate a maximum of 16 validators at a time. Yet, phragm\'{e}n works in a way that ensures that at most only one of the validators you voted for will become active, and only if the stake was high enough; since, validators are currently prioritized according to their total stake. \\

\noindent\textbf{Problems with participating in governance.}
The democracy and decentralization of the governance system is indeed questionable. The technical committee consists of Web3 Foundation and Parity Technologies, which are both founded and run by Gavin Wood, who also happens to be a ``prime voter" on the council. A prime voter is one who decides what the default vote is for any abstaining voters at the end of the voting period. In effect, Gavin Wood has direct influence on governance decisions; where as a council member, he holds the power to control the treasury and veto referenda, and in the technical committee, he provides the final say about the implementation of any referendum, even if it passed. To top that, the council is restricted to 13 members only. This brings fears that if only seven members are compromised, the integrity of the entire governance structure will collapse, such as in the nefarious Ronin Bridge attack \cite{ronin_bridge_attack}, where only five of nine validator keys were needed to succeed in a large-scale cryptocurrency heist worth around 624 millions USD.

Council members are also elected through phragm\'{e}n method just like validators; DOT holders back council candidates with their stake. At the time of writing, the minimum stake required to be elected as a council member is around 9.5 million\footnote{Data taken from Polkadot.js: \url{https://polkadot.js.org/apps/\#/council}, Accessed on April 2022} DOT, which is 171 million USD at \$18/DOT. Moreover, monetary punishment and restrictions make it unfavorable to participate in Polkadot's `democracy'. For instance, stake deposited by users is burned in such cases when a proposal is rejected. In addition, the reserved DOT for creating or endorsing a proposal can never be released until the proposal is brought to a vote (i.e., tabled), which is an indeterminate amount of time.   \\

\begin{table}
\centering
\caption{Polkadot's Parachain Winners in its First Batch of Auctions}
\label{tab:parachain_winners}
\resizebox{\linewidth}{!}{%
\begin{tabular}{|r|l|l|c|c|} 
\hline
\multicolumn{1}{|c|}{} & \textbf{Parachain} & \textbf{Role} & \begin{tabular}[c]{@{}c@{}}\textbf{Locked Stake }\\\textbf{(million DOT)}\end{tabular} & \begin{tabular}[c]{@{}c@{}}\textbf{Locked Stake }\\\textbf{(million USD)}\end{tabular} \\ 
\hline
1 & Acala & \begin{tabular}[c]{@{}l@{}}Polkadot-native \\Stablecoin hub\end{tabular} & 32 & 576~ \\ 
\hline
2 & Moonbeam & Smart contracts & 35 & 630~ \\ 
\hline
3 & Astar & Smart contracts & 10.3 & 185~ \\ 
\hline
4 & Parallel Finance & Assets transfer & 10.7 & 193~ \\
\hline
5 & Clover Finance & Smart contracts & 9.7 & 175~ \\
\hline
\end{tabular}}
\end{table}

\noindent\textbf{Problems with becoming a parachain.} Polkadot will support only 100 parachain slots in total, thus imposing a major limitation on true scalability of the system. Parachain slots are split between parachains and parathread pools, where a number of slots are reserved for common good parachains whose lease never expires, e.g., Statemint and public utility chains. Nonetheless, the cost to secure a parachain slot in the candle auctions is, simply put, almost prohibitive. As an example, the winning parachains of Polkadot's first batch of auctions\footnote{Data taken from Subscan: \url{https://polkadot.subscan.io/auction}} (concluded on December 17, 2021) and their stakes are listed in order in Table \ref{tab:parachain_winners}. 

The funds are reserved in an account created for the parachain during the entire lease period, up to 2 years. When the lease expires, the parachain has to enroll again in the auctions. Since slots are scarce, it is expected that the rising competition will increase the cost of securing a slot. If the parachain can no longer secure its slot, it has to simply either downgrade to a parathread or retire from the network altogether. In addition, any number of parathreads can exist in the parathread pool, but only a limited number can execute in each block. As a result, as the number of parathreads increases, smaller parathread projects will eventually not be able to include their blocks.  \\

\noindent\textbf{Transparency problems with the validators elections.} Polkadot uses a complex (computationally intensive) algorithm for proportionally justified representation in its elections. However, since the election is run by off-chain workers and the results are submitted later to the chain, the transparency of the results become questionable. Are the election results being validated by on-chain validators? What proof do the designated off-chain workers provide to attest the authenticity of their computations? The answers to these questions were not clear in the online documentations \cite{polkadotwiki} nor in the overview paper \cite{burdges2020overview}.

In addition, at the beginning of each era, a pool of 200 para-validators is formed from the larger validator set. It is said that the pool members are selected at random, and then groups of five are again formed randomly. It is not clear whether said randomness is provable using a VRF like BABE's block authoring assignments. By observing Polkadot.js, one of Polkadot's block explorers, we believe that these groups are seemingly rotated in a round-robin fashion, and the group assignments do not change during the epoch (4 hours). This brings up several security-related concerns. For example, can an adversary target a known majority in one of the para-validator groups to halt the operation of a parachain during an epoch? 

\begin{figure*}[htbp]
    \begin{minipage}[ct]{0.5\linewidth}
        \centering
        \includegraphics[scale=0.5]{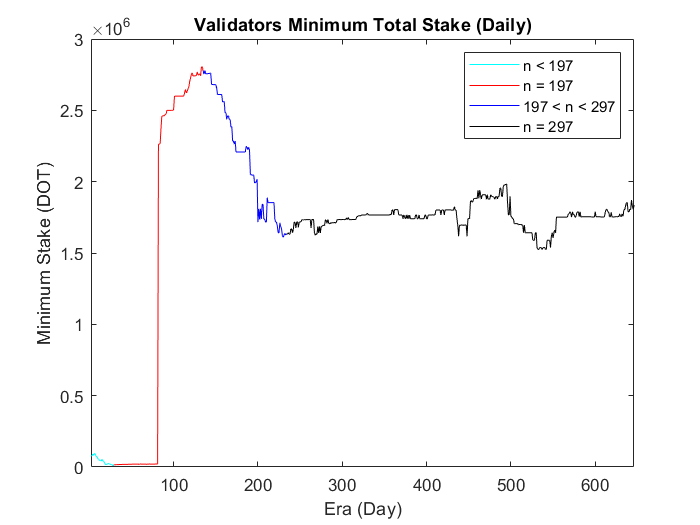}
        \caption{Validators Minimum Total Stake}
        \label{fig:validator_stake}
    \end{minipage} \hfill
     \begin{minipage}[ct]{0.5\linewidth}
        \centering
        \includegraphics[scale=0.53]{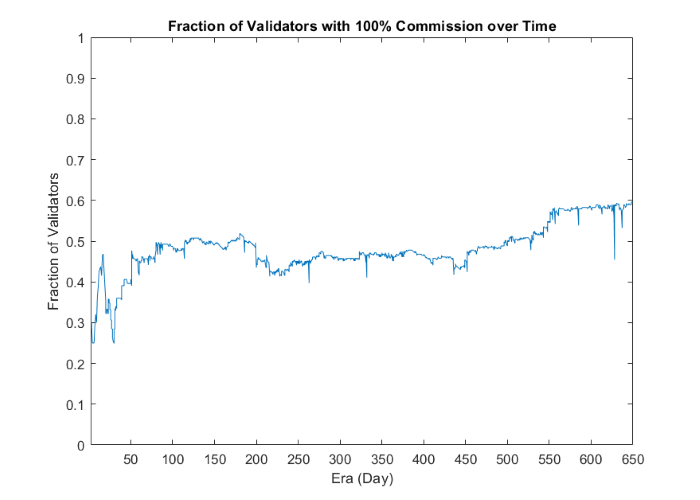}
        \caption{Validators with 100\% Commission Rate}
        \label{fig:validator_commission}
    \end{minipage}
\end{figure*}

\subsection{Empirical Analysis}
NPoS scheme is a crucial component of Polkadot that governs economic incentives and directly impacts overall security of the network. Therefore, we resorted to quantitative analysis to support the previously highlighted substantial limitations and contradictions. In what follows, we present our findings, primarily related to the nomination process, validators selection, and validator preferences. 

First, we had set up a Polkadot full archive node on an Ubuntu 20.04 LTS virtual environment. We then traced `NewSession' events emitted from June 2020 to April 2022 to obtain the corresponding validators elections data from the runtime storage. We parsed the data into a MySQL database (v8.0.28) using a Python (v3.10.2) script that we developed. From the data, we investigated the claims related to the validators minimum stake requirement and the validators commission rate. 

As shown in Figure \ref{fig:validator_stake}, the minimum stake required to enter the active set varies with the number of active validators ($n$). For instance, in the initial stages where $n < 197$, even though the size of the validators set was small, there was much less competition and traction in becoming a validator; and therefore, the required stake was negligible. The minimum stake declined as the number of validators increased. As of April 2022, the minimum amount required to become a validator is around 1.8 million DOT, and at approximately \$18 per DOT\footnote{CoinMarketCap: \url{https://coinmarketcap.com/currencies/polkadot-new/}, Accessed April 2022.}, that is a total of 32.4 million USD. It is unknown whether the minimum stake would further decrease as the number of validators progresses towards \numprint{1000}, because there are other factors affecting stake requirement, e.g., competition. 

In terms of validators' preferences, we find an alarming pattern. That is, the percentage of active validators that set 100\% commission is found to be increasing over time, where currently it has exceeded 60\%. Figure \ref{fig:validator_commission} depicts the fraction of active 100\%-commissioned validators over time and proves that a majority has been involved in this behavior. Typically, the perpetrators put forward the minimum self-stake and are backed by a few nominator accounts. Our hypothesis is that the nominator accounts also belong to the validator, otherwise there is no monetary incentive for backing a 100\%-commissioned validator that has almost no self-stake; risking painful slashing and also getting no rewards. We also hypothesize that this behavior is done to discourage other nominators from subscribing to the validator, so that the validator retains all the rewards for himself.

\subsection{Discussion}
Polkadot sets limits on various parameters (as shown in Table \ref{tab:scalability_limitations}) to balance between the performance and the security of the system. These bounds effectively restrict the actual scalability of the network. Even though the parameters can be modified later through governance, we notice that scalability is advertised as a prominent feature of Polkadot and a distinguishing factor from other blockchains. However, enforcing the mentioned limits, especially at such an early stage of the network's lifecycle, becomes a conceivable problem and makes us question whether 
the network can expand, as it continues to grow in size and features.

Polkadot claims that its NPoS economic incentive scheme provides higher inclusion for network participants than centralized systems \cite{polkadotnetwork}. Nevertheless, the limits induced on the staking system resulted in the presence of ``whales” and market exchange centers (e.g., Binance)---owners of a large amount of the network’s tokens, which is a typical phenomenon in traditional PoS systems. Due to the size of their holdings, these whales have the potential to unfairly influence the network and destabilize the system. Polkadot also claims that its NPoS system helps preventing the formation of validator pools, like what is seen in Bitcoin, since a larger number of nominators back a limited number of validators. However, this is not entirely true since majority of the validators charge full commission and hypothetically they are the ones staking DOT in their own favor. Validators who own large stake do not need nominators to back them up and tend to monopolize the rewards for themselves. In the future, this may lead to a centralization of power, contradicting with the concept of  decentralization envisioned by Web3 Foundation and described in Polkadot's whitepaper~\cite{wood2016polkadot}. 

Overall, we find that Polkadot is not so different from other PoS networks in terms of the ``rich gets richer" conundrum. The strict limits enforced on nominators, the prohibitive high stake required to become a validator, in addition to the seemingly not so democratic governance, eventually will lead to heavy centralization of power amongst the top few who own the most DOT. We believe that the contradictions and limitations discussed above call for further investigations and solutions.

\section{Conclusion and Future Work}
\label{sec:conclusion}
In this paper, we presented the first systematic study on the Polkadot ecosystem. To the best of our knowledge, we are the first ones to  discuss the complex architecture behind one of the most promising projects in the current DeFi landscape, its protocols, and economic model. We also provided our preliminary insights about the analyzed architecture, including various contradictions regarding how Polkadot is presented to the general public and the way it is implemented. 
For instance, according to our results, Polkadot shows several limitations, at least in its current implementation, that can lead to heavy centralization of power, excluding regular users from network maintenance and governance. 
In fact, the constraints on the number of validators and nominators, together with the NPoS consensus in use, restrict the participation in the protocol to a few entities with considerable funds (around 32 Million USD to become a validator), leaving no room for ordinary users.

Our results call for a more in-depth analysis of the on-chain data to further investigate the real potential and functioning of the multi-chain protocols behind the  multi-billion Polkadot project, as well as its systemic weaknesses. 

\section*{Acknowledgment}
This publication was made possible by an award [GSRA7-2-0527-20101] and [NPRP-S-11-0109-180242] from Qatar National Research Fund (a member of Qatar Foundation). The contents herein are solely the responsibility of the author.

\balance
\bibliographystyle{IEEEtran}
\bibliography{main}

\begin{thebibliography}{10}
\providecommand{\url}[1]{#1}
\csname url@samestyle\endcsname
\providecommand{\newblock}{\relax}
\providecommand{\bibinfo}[2]{#2}
\providecommand{\BIBentrySTDinterwordspacing}{\spaceskip=0pt\relax}
\providecommand{\BIBentryALTinterwordstretchfactor}{4}
\providecommand{\BIBentryALTinterwordspacing}{\spaceskip=\fontdimen2\font plus
\BIBentryALTinterwordstretchfactor\fontdimen3\font minus
  \fontdimen4\font\relax}
\providecommand{\BIBforeignlanguage}[2]{{%
\expandafter\ifx\csname l@#1\endcsname\relax
\typeout{** WARNING: IEEEtran.bst: No hyphenation pattern has been}%
\typeout{** loaded for the language `#1'. Using the pattern for}%
\typeout{** the default language instead.}%
\else
\language=\csname l@#1\endcsname
\fi
#2}}
\providecommand{\BIBdecl}{\relax}
\BIBdecl

\bibitem{guo2019graph}
D.~Guo, J.~Dong, and K.~Wang, ``Graph structure and statistical properties of
  ethereum transaction relationships,'' \emph{Information Sciences}, vol. 492,
  pp. 58--71, 2019.

\bibitem{coinmarketcap}
``Coin market cap,'' \url{https://coinmarketcap.com/}, accessed: 2022-04-20.

\bibitem{wood2016polkadot}
G.~Wood, ``Polkadot: Vision for a heterogeneous multi-chain framework,''
  \emph{White Paper}, vol.~21, pp. 2327--4662, 2016.

\bibitem{polkadotnetwork}
``Polkadot network,'' \url{https://polkadot.network/}, accessed: 2022-01-30.

\bibitem{9070689}
A.~Ismailisufi, T.~Popović, N.~Gligorić, S.~Radonjic, and S.~Šandi, ``A
  private blockchain implementation using multichain open source platform,'' in
  \emph{2020 24th International Conference on Information Technology (IT)},
  2020, pp. 1--4.

\bibitem{back2014enabling}
A.~Back, M.~Corallo, L.~Dashjr, M.~Friedenbach, G.~Maxwell, A.~Miller,
  A.~Poelstra, J.~Tim{\'o}n, and P.~Wuille, ``Enabling blockchain innovations
  with pegged sidechains,'' \emph{URL: http://www. opensciencereview.
  com/papers/123/enablingblockchain-innovations-with-pegged-sidechains},
  vol.~72, 2014.

\bibitem{10.1007/978-3-030-43725-1_3}
A.~Kiayias and D.~Zindros, ``Proof-of-work sidechains,'' in \emph{Financial
  Cryptography and Data Security}, A.~Bracciali, J.~Clark, F.~Pintore, P.~B.
  R{\o}nne, and M.~Sala, Eds.\hskip 1em plus 0.5em minus 0.4em\relax Cham:
  Springer International Publishing, 2020, pp. 21--34.

\bibitem{8431965}
L.~Kan, Y.~Wei, A.~Hafiz~Muhammad, W.~Siyuan, L.~C. Gao, and H.~Kai, ``A
  multiple blockchains architecture on inter-blockchain communication,'' in
  \emph{2018 IEEE International Conference on Software Quality, Reliability and
  Security Companion (QRS-C)}, 2018, pp. 139--145.

\bibitem{belchior2021survey}
R.~Belchior, A.~Vasconcelos, S.~Guerreiro, and M.~Correia, ``A survey on
  blockchain interoperability: Past, present, and future trends,'' \emph{ACM
  Computing Surveys (CSUR)}, vol.~54, no.~8, pp. 1--41, 2021.

\bibitem{9154979}
A.~Alkhodair, S.~Mohanty, E.~Kougianos, and D.~Puthal, ``Mcpora: A multi-chain
  proof of rapid authentication for post-blockchain based security in large
  scale complex cyber-physical systems,'' in \emph{2020 IEEE Computer Society
  Annual Symposium on VLSI (ISVLSI)}, 2020, pp. 446--451.

\bibitem{8761448}
A.~Ahmad, M.~Saad, L.~Njilla, C.~Kamhoua, M.~Bassiouni, and A.~Mohaisen,
  ``Blocktrail: A scalable multichain solution for blockchain-based audit
  trails,'' in \emph{ICC 2019 - 2019 IEEE International Conference on
  Communications (ICC)}, 2019, pp. 1--6.

\bibitem{cosmos}
``Cosmos whitepaper,'' \url{https://v1.cosmos.network/resources/whitepaper},
  accessed: 2022-04-24.

\bibitem{polkadotwiki}
``Polkadot wiki,'' \url{https://wiki.polkadot.network/docs/}, accessed:
  2022-01-30.

\bibitem{cevallos2021verifiably}
A.~Cevallos and A.~Stewart, ``A verifiably secure and proportional committee
  election rule,'' in \emph{Proceedings of the 3rd ACM Conference on Advances
  in Financial Technologies}, 2021, pp. 29--42.

\bibitem{burdges2020overview}
J.~Burdges, A.~Cevallos, P.~Czaban, R.~Habermeier, S.~Hosseini, F.~Lama, H.~K.
  Alper, X.~Luo, F.~Shirazi, A.~Stewart \emph{et~al.}, ``Overview of polkadot
  and its design considerations,'' \emph{arXiv preprint arXiv:2005.13456},
  2020.

\bibitem{babe_web3}
H.~K. Alper, ``Babe,''
  \url{https://research.web3.foundation/en/latest/polkadot/block-production/Babe.html},
  accessed: 2022-03-8.

\bibitem{stewart2020grandpa}
A.~Stewart and E.~Kokoris-Kogia, ``Grandpa: a byzantine finality gadget,''
  \emph{arXiv preprint arXiv:2007.01560}, 2020.

\bibitem{polkadot_postmortem}
B.~Köcher, ``A polkadot postmortem - 24.05.2021,''
  \url{https://polkadot.network/blog/a-polkadot-postmortem-24-05-2021/},
  accessed: 2022-03-12.

\bibitem{ronin_bridge_attack}
``Ronin attack shows cross-chain crypto is a ‘bridge’ too far,'' \url{
  https://www.coindesk.com/layer2/2022/04/05/ronin-attack-shows-cross-chain-crypto-is-a-bridge-too-far/},
  accessed: 2022-04-26.

\end{thebibliography}

\end{document}